\documentclass{article}
\usepackage{amsmath,amsthm}
\usepackage{amsfonts}
\usepackage{amssymb}

\usepackage{url}
\usepackage{hyperref}

\theoremstyle{plain}
\newtheorem{theorem}{Theorem}
\theoremstyle{definition}

\newtheorem{remark}[theorem]{Remark}

\newcommand{\C}{\mathbb{Q}}

\begin{document}

\title{Submodule approach to creative telescoping \\[10pt]
\Large In memory of Marko Petkov\v{s}ek}
\author{Mark van Hoeij\footnote{Supported by NSF grant 2007959.}}
\maketitle

\begin{abstract}
This paper proposes ideas to speed up the process of creative telescoping, particularly when the telescoper is reducible.
One can interpret telescoping as computing an annihilator $L \in D$ for an element $m$ in a $D$-module $M$.
The main idea in this paper is to look for submodules of $M$. If $N$ is a non-trivial submodule of $M$,
constructing the minimal annihilator $R$ of the image of $m$ in $M/N$ gives a right-factor of $L$ in $D$.
Then $L = L' R$ where the left-factor $L'$ is the telescoper of $R(m) \in N$.
To expedite computing $L'$, compute the action of $D$ on a natural basis of $N$, then obtain $L'$
with a cyclic vector computation.

The next main idea is to construct submodules from automorphisms, if we can find some.
An automorphism with distinct eigenvalues
can be used to decompose $N$ as a direct sum $N_1 \oplus \cdots \oplus N_k$.  Then $L'$ is the LCLM (Least Common Left Multiple)
of $L_1, \ldots, L_k$ where $L_i$ is the telescoper of the projection of $R(m)$ on $N_i$. An LCLM can greatly increase the degrees
of coefficients, so $L'$ and $L$ can be much larger expressions than the factors $L_1,\ldots,L_k$ and $R$.
Examples show that computing each factor $L_i$ and $R$ separately can save a lot of CPU time
compared to computing $L$ in expanded form with
standard creative telescoping.
\end{abstract}

\section{Introduction}
A common way to find a closed form expression for a sequence $a_n$ or its generating function $y(x) = \sum a_n x^n$
is to {\em guess an equation}  (a {\em recurrence} for $a_n$ or a {\em differential equation} for $y(x)$,  if it exists) and then solve it.
%
%
To prove correctness, the main technique is creative telescoping, which is highly successful for proving identities \cite{AB}.

One can interpret telescoping as computing an annihilator (called {\em telescoper}) of some $m$ in some $D$-module $M$, designed
in such a way that annihilators of $m$ will also\footnote{The implication is in only one direction, 
so a minimal telescoper need not be a minimal recurrence/equation.}
 annihilate $a_n$ or $y(x)$.
In telescoping, a {\em certificate} is a proof that $L$ annihilates $m$, which after some checks implies that $L$ is a provably correct equation. 
The OEIS (Online Encyclopedia for Integer Sequences) contains many examples where guessed equations or formulas for $a_n$ or $y(x)$
were proven with creative telescoping. \\

\noindent {\bf Notation:} Let ${\rm Ann}_D(m, M) := \{ L \in D \ | \ L(m){\rm \ vanishes \ in \ }M\}$  denote the annihilator of $m$ in a $D$-module $M$.
If $m \in M$ and $N$ is a submodule of $M$, then ${\rm Ann}_D(m, M/N) = \{ L \in D \ | \ L(m) \in N\}$. 
If the left-ideal ${\rm Ann}_D(m, M/N) \subseteq D$ is principal and not zero, then ${\rm Ann}^{\rm min}_D(m, M/N)$ denotes a generator, called {\em minimal annihilator of $m$ in $M/N$}.
For uniqueness, we will assume it to be monic.

\begin{remark} \label{Remark1}
${\rm Ann}_D(m, M) \subseteq {\rm Ann}_D(m, M/N)$ so 
the minimal annihilator of $m$ in $M/N$ is a {\em right-factor} of the minimal annihilator of $m$ in $M$.
\end{remark}
The remark is obvious, clearly if $L(m)$ vanishes in $M$, then it also vanishes in $M/N$.
But it is very useful. It helps to explain why some telescopers must be reducible, why a minimal telescoper need not be
a minimal recurrence, and how to find factors if suitable submodules are known. We will briefly summarize this,
and then illustrate it with examples.
To efficiently compute $L := {\rm Ann}^{\rm min}_D(m, M)$,
\begin{enumerate}
	\item  
	The goal in Sections~\ref{Section3} and~\ref{sym} is to find
	submodule(s) $N \subsetneq M$. Then $R := {\rm Ann}^{\rm min}_D(m, M/N)$
	is a {\em right-factor} of $L$, and $L = L' R$ where the {\em left-factor} $L' := {\rm Ann}^{\rm min}_D( R(m), N)$ is the telescoper of $R(m)$.
	By computing the action of $D$ on a basis of $N$ and  
	we can obtain $L'$ through a cyclic vector computation (Section~\ref{Section4}).
	\item Section~\ref{sym} illustrates using automorphisms to decompose a $D$-module as a direct sum.
	Computing an annihilator $L_1,L_2,\ldots$ for each component can lead to
	a compact {\em LCLM representation}. 
\end{enumerate}
The author thanks Shaoshi Chen, Alin Bostan, Manuel Kauers, Christoph Koutschan,  and Fr\'ed\'eric Chyzak for discussions on telescoping, and the referees.
The author also acknowledges support of the Institut Henri Poincar\'e (UAR 839 CNRS-Sorbonne Universit\'e), and LabEx CARMIN (ANR-10-LABX-59-01).

\section{Creative telescoping for hypergeometric terms} \label{Section2}
Creative telescoping works in many contexts, we pick one to illustrate the ideas.
An expression $H = H(n,k)$ is a  {\em hypergeometric term}
if both $R_1 := S_n(H)/H$ and $R_2 := S_k(H)/H$ are non-zero elements of $\C(n,k)$.
Here $S_n$ resp. $S_k$ are the {\em shift operators} that send an expression $f(n,k)$ to $f(n+1,k)$ resp. $f(n,k+1)$.

Gosper's algorithm \cite{Gosper} decides if $H = \Delta_k( c H )$ for some $c \in \C(n,k)$, where the
{\em difference operator} $\Delta_k = S_k - 1$ sends $f(n,k)$ to $f(n,k+1) - f(n,k)$.
Zeilberger's insight \cite{ZeilbergerAlgorithm}, {\em creative telescoping}, was that Gosper's algorithm could be modified to compute
an $L \in \C(n)[S_n]$ with $L(H) = \Delta_k( c H )$
for some $c \in \C(n,k)$.  See \cite{when} for when such {\em telescoper} $L$ and {\em certificate} $c$ must exist.
Since $\sum_{k \in \mathbb{Z}} \Delta_k( f ) = 0$ when $f(k) \neq 0$ for only finitely $k  \in \mathbb{Z}$,
creative telescoping typically produces a provably correct recurrence operator $L \in \C(n)[S_n]$  (after some checks such as non-zero denominators)
for sequences of the form
\begin{equation}
	a(n) := \sum_{k \in \mathbb{Z}} H(n,k).  \label{an}
\end{equation}

The statement that $S_n(H)/H$ and $S_k(H)/H$ are both in $\C(n,k) - \{0\}$ means that
\begin{equation}
	\Omega := \C(n,k) \cdot H \label{Omega}
\end{equation}
is a $D_{n,k}$-module where $D_{n,k} := \C(n,k)[S_k, S_k^{-1}, S_n, S_n^{-1}]$. 

A telescoper for $H$ is a non-zero operator $L \in D  := \C(n)[S_n, S_n^{-1}]$ for which $L(H) = \Delta_k( c H )$ for some $c \in \C(n,k)$.
This condition can be reformulated as saying
that $L(m) = 0$ where
\begin{equation}
	m := H + \Delta_k( \Omega )   \label{m}
\end{equation}
is the image of $H$ in
\[
	M := \Omega / \Delta_k( \Omega ).
\]
This $M$ is a $D$-module. Working modulo $\Delta_k(\Omega)$ causes $S_k$ to act as the identity on $M$.
However, $M$ is not a $D_{n,k}$-module 
because $\Delta_k( \Omega )$ is not a $\C(k)$-vector space.

An annihilator (minimal or not) of $m$ is called a {\em telescoper}. 
Saying that $L$ annihilates $m$ is equivalent to saying that $L(H) \in \Delta_k(\Omega)$, in other words $L(H) = \Delta_k( c H )$ for
some $c \in \C(n,k)$, the {\em certificate}. 
Reduction-based telescoping algorithms find $L$ (i.e. a $\C(n)$-linear relation between $m, S_n(m), S_n^2(m), \ldots$)
by {\em reducing} 
$H, S_n(H), S_n^2(H), \ldots$ 
mod $\Delta_k( \Omega )$.
Those reductions allow us to compute in the $D$-module $M$.
Our focus will be submodules of $M$.

\section{Exploiting submodules} \label{Section3}

Gosper's algorithm provides a zero test for $M = \Omega / \Delta_k(\Omega)$. The more general problem
of {reducing} elements of $\Omega$ to a standard form modulo $\Delta_k(\Omega)$ is solved
in {\em reduction-based}\hspace{1pt} telescoping algorithms such as \cite{StandardForm1, PolynomialReduction}.
This makes it possible to compute in $M$.

The main idea in this paper is to work in $M/N$, which means that additional reductions are allowed, namely modulo $N$. 
These additional reductions might (Question~\ref{q5} in Section~\ref{FutureResearch}) help to speed up the reduction process, but the main
benefit is that they could lead to finding a right-factor instead of the full telescoper, see Remark~\ref{Remark1}.
Of course, this benefit is not always achievable (a telescoper can be irreducible).

Our main example, sent to us by Shaoshi Chen, is
\begin{equation}
	\label{example1}
	H = \frac{ \left( \hspace{-5pt} \begin{array}{c} n \\ k \end{array} \hspace{-5pt} \right)^{\hspace{-3pt} 7}}{ 2n + 3k }.
\end{equation}
This defines a sequence
\[
	a(n) := \sum_{k=0}^n \frac{ \left( \hspace{-5pt} \begin{array}{c} n \\ k \end{array} \hspace{-5pt} \right)^{\hspace{-3pt} 7}}{ 2n + 3k },  \ \ \ \ n = 1,2,3,\ldots
\]
and the goal is to obtain a provably correct recurrence operator $L \in D$ for $a(n)$  
by computing an annihilator for~(\ref{m}).
We can express $H$ in terms of a simpler hypergeometric term $H_0$ as follows: $H = R_0 H_0$ where
\[
	R_0 := \frac{1}{2n + 3k} \ \ \ {\rm and} \ \ \  H_0 := \left( \hspace{-5pt} \begin{array}{c} n \\ k \end{array} \hspace{-5pt} \right)^{\hspace{-3pt} 7}\hspace{-2pt}.\]
The simplest way to express $\Omega = \C(n,k) H$ is to write it as
\begin{equation} \label{simpler}
	\Omega = \C(n,k) H_0.
\end{equation}
This is the same module, but now elements will be expressed in terms of $H_0$.

In our example, the simplified representation of $\Omega$ in~(\ref{simpler}) is obvious, but otherwise we can use  \cite{AP2}.
(Brief summary: factor $S_k(H)/H = c \prod f_i^{e_i}$ with $c \in \C(n)$, $e_i \in \mathbb{Z}$, and $f_i$ monic and irreducible in $\C(n)[k]$.
As long as this product contains {\em shift-equivalent} factors, $f_i = S_k^m(f_j)$ for some non-zero $m \in \mathbb{Z}$, replace one of them with the other.
Then take $H_0$ for which $S_k(H_0)/H_0$ equals the reduced product.)

Let $N = \{ P H_0 + \Delta_k(\Omega) \ | \ P \in \C(n)[k]\}$,
the $\C(n)$-subspace of $M$ spanned by
\[
	m_d := k^d H_0 + \Delta_k(\Omega)
\]
for $d \geq 0$.
Reduction techniques from modern telescoping algorithms (Hermite reduction \cite{StandardForm1}, shell reduction and polynomial  reduction \cite{PolynomialReduction})
show that $N$ is a $D$-module with $\C(n)$-basis $\{m_0,\ldots,m_6\}$, as follows. 
The actions of $S_n$ and $S_n^{-1}$ on $H_0$ are given by
\[
	R_1 := \frac{S_n(H_0)}{H_0} = \left( \frac{n+1}{n-k+1}  \right)^7 \ \ \ {\rm and} \ \ \ \tilde{R}_1 := \frac{S_n^{-1}(H_0)}{H_0}  = \left( \frac{n-k}{n}  \right)^7.
\]
Now  $S_n(m_d) = k^d R_1 H_0 + \Delta_k(\Omega)$. The claim that $N$ is a $D$-module means that this should be in $N$.
That is not obvious because of the denominator $(n-k+1)^7$ in $R_1$.  
The denominator cancels by {\em reducing} mod $\Delta_k(\Omega)$. The set $\Delta_k(\Omega)$ is a $D$-module and is closed under $S_k$ and $S_k^{-1}$,
whose actions are described by
\[
	R_2 := \frac{S_k(H_0)}{H_0} = \left( \frac{n-k}{k+1}  \right)^7 \ \ \ {\rm and} \ \ \ \tilde{R}_2 := \frac{S_k^{-1}(H_0)}{H_0}  = \left( \frac{k}{n-k+1}  \right)^7.
\]
We use that to write an ansatz $G \in \Delta_k(\Omega)$ such that subtracting $G$ from $k^d R_1 H_0$
and solving $\C(n)$-linear equations cancels the denominator.
That writes $S_n( m_d )$ as $P H_0 + \Delta_k(\Omega)$ for some {\em polynomial} $P \in \C(n)[k]$.

The next step is to reduce the degree of $P$.
We can find polynomials $P_d \in \C(n)[k]$ of degree $d$ with $P_d H_0 \in \Delta_k(\Omega)$ for every $d \geq 7$. Subtracting a $\C(n)$-linear combination
reduces the degree of $P$ to $\leq 6$, thereby writing $S_n(m_d)$ as a unique linear combination of $\{m_0,\ldots,m_6\}$.
An accompanying computer file \cite{website} performs these reductions for this example, but it is known
how to do this more generally, see reduction based telescoping papers such as~\cite{StandardForm1, PolynomialReduction}.

What makes the $D$-submodule $N \subseteq M$ useful for Remark~\ref{Remark1}.
is the fact that $m = H + \Delta_k(\Omega) $  
is {\em not} in $N$, in other words,
$\overline{m} := m+ N$ is not zero in $M/N$. Our goal is now to compute its annihilator.

Since $N$ contains $P H_0 + \Delta_k(\Omega)$ for any polynomial $P \in \C(n)[k]$,
we can annihilate $\overline{m} \in M/N$ by eliminating $k$ from the denominator.
The denominator $2n+3k$ of $H$ in~(\ref{example1})  can be interpreted (just remove the factor $H_0$) as a pole of order 1 at $k = -2n/3$.
Now $S_n^3(S_k^{-2}(H))$ has the same denominator $2(n+3) + 3(k-2) = 2n + 3k$.
Substituting $k = -2n/3$ in the quotient $S_n^3(S_k^{-2}(H)) / H$ gives
\[
	r := \left(\frac{54(n+2)(n+1)n(2n+3)}{5(5n+12)(5n+9)(5n+6)(5n+3)}\right)^7.
\]
Then $S_n^3(S_k^{-2}(H))$ has the same pole and residue at $k = -2n/3$ as $r H$, so the denominator in
\[
	S_n^3(S_k^{-2}(H)) - r H
\]
cancels out. Hence $S_n^3(S_k^{-2}(m)) - r m \in N$.
Now $S_k^{-2}(m) = m$ because $S_k$ is the identity on $M$. After all, computing in $M$ means working modulo $\Delta_k( \Omega )$ and $\Delta_k = S_k - 1$.
It follows that $R(m) \in N$, in other words $R(\overline{m}) = 0$
where
\[
	R = S_n^3  - r \in D. 
\]
The denominators of $S_n^i(H)$ for $0 \leq i < 3$ are  $2n + 3k$, $2(n+1)+3k$ and $2(n+2)+3k$. These do not cancel modulo $\Delta_k(\Omega)$,
so $R$ is minimal.
So
$L = {\rm Ann}_D^{\rm min}(m, M)$
must have $R = {\rm Ann}_D^{\rm min}(m, M/N)$ as right-factor. We multiply $R$ by its denominator to reduce it to an element of $\C[n, S_n]$.
The next task is to compute the left factor; a telescoper for $R(m) \in N$.

\begin{remark} If the denominator contains more factors then we can apply the
above process for each $k$-shift-equivalence class of factors.
Say there are $m$ such equivalence classes $C_1,\ldots,C_m$.
Then the submodule $N_i$ should allow all denominators, except those in $C_i$.
This then produces a telescoper $R_i$ that eliminates the denominators in $C_i$.
Then $R := {\rm LCLM}(R_1,\ldots,R_m)$ should send $m$ to an element of $N$,
after which we can proceed as in Section~\ref{Section4} below.
\end{remark}
 
 \begin{remark} Obtaining a right-factor of the telescoper by considering denominators already appeared in \cite{pretelescoper} in the differential case.
But the submodule idea is more general, 
as illustrated in Section~\ref{sym}.
 \end{remark}
 
\section{Using a basis of $N$} \label{Section4}
In the main example, $N$ is a $\C(n)$-vector space with basis $B := \{m_0,\ldots,m_6\}$.
However, writing $R(m)$ as a $\C(n)$-linear combination of $B$, the coefficients in $\C(n)$ are surprisingly large.
Instead of repeatedly applying $S_n$ to this large expression, each time reducing mod $\Delta_k(\Omega)$,
it is more efficient to reduce $S_n(m_d)$ for $d = 0,\ldots,6$ and obtain a $7 \times 7$ matrix for the action of $S_n$ on $B$.

After this pre-computation, no more reductions mod $\Delta_k(\Omega)$ are needed to compute $S_n^i(R(m))$ for $i = 0,\ldots,7$,
just substitutions $n \mapsto n+1$ and matrix-vector products.  Finally, by computing a $\C(n)$-linear relation between the $S_n^i(R(m))$,
an annihilator $L'$ for $R(m)$ is found of order $7 = {\rm dim}_{\C(n)}(N)$. (This equality implies that $R(m)$ generates the $D$-module $N$, it is a {\em cyclic vector}.)

The product $L' R \in \C(n)[S_n]$ is the {\em minimal telescoper} for $H$. It equals the telescoper found by Maple's SumTools[Hypergeometric][Zeilberger] program.
But it is not the {\em minimal recurrence} for sequence $(\ref{an})$.
To understand why, note that $N$ has obvious
submodules, which is the topic of the next section.

\section{Automorphisms} \label{sym}
The function $H_0(n,k)$ equals $H_0(n,n-k)$, so the modules $M$ and $N$
have an automorphism $\phi$ that sends $k$ to $n-k$.  Then $\phi^2 = 1$ so we can decompose  $N = N_+ \oplus N_-$ 
where $N_+$ and $N_-$ are the eigenspaces of $\phi$ with eigenvalues $\pm 1$.
The projections to $N_+$ and $N_-$ are $\phi_+ = (1 + \phi)/2$ and $\phi_- = (1 - \phi)/2$.

Since $H(n,k)$ and $H(n,n-k)$ take the same values for  $k \in \{0,\ldots,n\}$ (just not in the same order)
we find that  $a(n) =  \sum_{k=0}^n H$ equals  $\sum_{k=0}^n \phi_+(H)$,  and  that  $\sum_{k=0}^n \phi_-(H)$ equals zero.
A {\em recurrence} can ignore the $N_-$ component since it adds 0 to $a(n)$.
But a {\em telescoper} annihilates $m$, so $L'$ must annihilate {\em both} $N_+$ and $N_-$ components of $R(m)$.
That causes the minimal telescoper $L' R$ to be larger than the minimal recurrence.

The $\phi$-invariant subring of $\C(n)[k]$ is $\C(n)[ k(n-k) ]$.
A basis of $N_+$ is then $B_+ := \{ (k(n-k))^i  \cdot H_0 + \Delta_k(\Omega) \ | \ 0 \leq i \leq 3\}$.
Although $N_-$ is not needed for a recurrence,
for now we include it anyway so that our telescoper matches the standard one.
A basis of $N_-$ is $\{ (2k-n) \cdot (k(n-k))^i \cdot H_0 + \Delta_k(\Omega)  \ | \ 0 \leq i \leq 2\}$ (the factor $2k-n = k - \phi(k)$ has eigenvalue $-1$).

We can pre-compute a 4 by 4 matrix over $\C(n)$ that represents the action of $S_n$ on $B_+$.
Then write $\phi_+(R(m))$ as a linear combination of this basis, and apply linear algebra
to obtain its annihilator $L_4$ which has order 4.  Then $L_4 R$ is a recurrence for $a(n)$ of order~7, which is minimal\footnote{Once any recurrence is found, then with initial conditions
my implementation MinimalRecurrence in Maple's LREtools package computes the minimal recurrence. But ideally one would try to avoid computing a non-minimal recurrence, see also
Question~\ref{item3} in Section~\ref{FutureResearch}.}.

Doing the same for $N_-$, we quickly find the annihilator $L_3$ of order 3 for the $N_-$ component
of $R(m)$.  Then $L' = {\rm LCLM}(L_4, L_3)$ annihilates $R(m)$ 
and the product $L := L' \cdot R$ is the minimal telescoper for $H$.

\begin{remark}[Size]
Taking the LCLM of $L_4$ and $L_3$ significantly increases the degree in $n$, and with it, the overall expression size.
The expression for $L$ is much smaller in factored form  ${\rm LCLM}(L_4, L_3) \cdot R$ than in expanded form.
Our approach of using submodules to directly computing individual factors of $L$ has several advantages. Factoring is an important
step towards solving. So the telescoper is more
useful in factored form, and its smaller\footnote{A minimal telescoper $L$ often has multiples 
that are smaller than $L$ in terms of degree or bit-size~\cite{curve}
despite having higher order.
However, the author has not encountered examples where a minimal telescoper was smaller than its right-factor(s).}
 size speeds up computations.
The factored form also encodes the minimal recurrence in a convenient way, simply toss $L_3$ to obtain the minimal recurrence $L_4 R$.
\end{remark}

\begin{remark}[Timing]
The example runs in about 0.89 seconds on an Apple M1 Pro, which is 
much faster than existing implementations.
Our Maple file \cite{website} for this example performs reductions mod $\Delta_k( \Omega )$ in an ad-hoc manner;
a proper comparison should use a more systematic implementation for reduction, finding and exploiting
submodules and automorphisms, etc.
We expect the efficiency advantage to remain, 
simply because the combined bitsize
of $L_4, L_3$, and $R$ is more than 6 times smaller than the bitsize of $L$.
\end{remark}

\begin{remark} \label{s} Let $s$ be a positive integer,
\[ H :=  \left( \hspace{-5pt} \begin{array}{c} n \\ k \end{array} \hspace{-5pt} \right)^{\hspace{-3pt} s}   \ \ \ {\rm and}  \ \ \  r := \lfloor \frac{s+1}2 \rfloor. \]
Let $m_d$ be the image of $k^d H$ in $M = \Omega/\Delta_k(\Omega)$ with $\Omega = \C(n,k) H$.
Like in Section~\ref{sym}, reduction shows that $\{ m_d \ | \ 0 \leq d \leq 2r-2 \}$ is a basis of the $D$-module $N$ generated by $\{ m_d \ | \ d \geq 0\}$,
$N_+$ has basis $\{ (k(n-k))^d  \cdot H + \Delta_k(\Omega) \ | \ 0 \leq d \leq r-1\}$,
and $N_-$ has basis $\{ (2k-n) \cdot  (k(n-k))^d  \cdot H + \Delta_k(\Omega) \ | \ 0 \leq d \leq r-2\}$.
Since the image of $H$ in $M$ is $m_0 \in N_+$, the telescoper $L$ of $H$
must have ${\rm order}(L) =  {\rm dim}_{\C(n)}(D\,m_0) 
\leq {\rm dim}_{\C(n)}(N_+) = r$. A theorem by Straub and Zudilin \cite[Theorem~1.1]{StraubZudilin} says
${\rm order}(L) \geq r$ and hence this must be an equality.
\end{remark}

\section{An example with more automorphisms} \label{more}
Let $H :=   \left( \hspace{-5pt} \begin{array}{c} 3n \\ 3k \end{array} \hspace{-5pt} \right)^{\hspace{-3pt} 2}   \left( \hspace{-5pt} \begin{array}{c} 3n \\ 3k+1  \end{array}  \hspace{-5pt} \right)$, 
$\Omega := \C(n,k) H$,  and $M := \Omega / \Delta_k(\Omega)$.
Let $\phi$ send $f(n,k)$ to $f(n, n-k)$ as before, and let $\tau$ send $f(n,k)$ to $f(n, k+1/3)$.
Then $\phi(H)/H$ and $\tau(H)/H$ are in $\C(n,k)$ so $\phi$ and $\tau$ act on $\Omega$. These
$\phi$ and $\tau$ are also $D$-module automorphisms of $M$, and
$M = M_+ \oplus M_-$ where $M_+$ and $M_-$ are the eigenspaces of $\phi$ for the eigenvalues $\pm 1$.

Now $\tau^3 = S_k$ acts trivially on $M$, so $\tau^3 - 1 = (\tau - 1)(\tau^2 + \tau + 1)$ is zero on $M$. We have two options:
\begin{enumerate}
\item Write $M$ as a direct sum $M = M_{\tau -1} \oplus M_{\tau^2 + \tau + 1}$\,, the kernels of $\tau-1$ and $\tau^2 + \tau + 1$ respectively.
\item Extend the constants to $\C(\omega)$ and write $M = M_{\tau -1} \oplus M_{\tau-\omega} \oplus M_{\tau-\omega^2}$\,, the eigenspaces of $\tau$
for the eigenvalues $1, \omega, \omega^2$ where $\omega^2 + \omega + 1 = 0$.
\end{enumerate}

One might expect to find 6 non-zero submodules of $M$ if we intersect $M_+$ and $M_-$ with
the three submodules in option 2.
However, this is not the case because $\phi$ is not an automorphism
of $M_{\tau-\omega}$. Instead, it is an isomorphism\footnote{The observation $M_{\tau-\omega} \cong M_{\tau-\omega^2}$
explains an observation that the telescoper of $H$ has two factors $L_2,L_4$ with $D/DL_2 \cong D/DL_4$.}
from $M_{\tau-\omega}$ to $M_{\tau-\omega^2}$
and so $M_+ \bigcap M_{\tau-\omega} = \{0\}$.

So we choose option 1 instead.  Intersecting the two $M_{\pm}$ with $M_{\tau-1}$ and $M_{\tau^2+\tau+1}$ 
decomposes $M$ as a direct sum of 4 non-zero submodules.
We can compute the 4 components of $H$ in these 4 submodules, and then compute the telescoper of each component.

The easiest way to do this is as follows. The projections $(1+\phi)/2$ and $(1-\phi)/2$ from Section~\ref{sym}
produce the images $H_{\pm}$ of $H$ in $M_{\pm}$.
The projections from $M$  to $M_{\tau-1}$ and $M_{\tau^2+\tau+1}$ are
$(1+ \tau + \tau^2)/3$ and  $(2-\tau-\tau^2)/3$.  Applying these to $H_+$ and $H_-$ gives the 4 components of $H$.

We computed telescopers $L_1,L_2,L_3, L_4 \in \C(n)[S_n]$ for each component.
Each turned out to be irreducible. The orders are $2, 3, 1, 3$. The telescoper of $H$ is ${\rm LCLM}(L_1,L_2,L_3,L_4)$ and has order $9$.
We get the minimal recurrence $L_{\rm min} = {\rm LCLM}(L_1,L_2)$ of order 5 for $a(n) := \sum_{k} H$ simply by omitting the telescopers $L_3,L_4$ of the projections of $H_-$ on $M_{\tau-1}$ and $M_{\tau^2+\tau+1}$.

Alternatively, we could compute bases of submodules like in Sections~\ref{Section4} and~\ref{sym}.
For computational details see the accompanying website~\cite{website}.

\section{Research questions} \label{FutureResearch}
\begin{enumerate}
\item By {\em invariant data} of $L$, we mean data
such as the characteristic polynomial, or the $p$-curvature, that is the same for any $\tilde{L}$ with $D/D \tilde{L} \cong D/D {L}$.
The question is, if we know the module $D/DL$ up to isomorphism, but have not computed $L$, then how do we compute
invariant data directly from the module, without computing $L$ itself?

The characteristic polynomial for 
$L$ from Remark~\ref{s}
appears to have the following roots $\{ (\zeta + \zeta^{-1})^s \, | \, \zeta^s = 1, \zeta \neq -1, \zeta^2 \neq -1 \}$.
The question is, how to compute such data directly from the module $N_+ \cong D/DL$  
without having to compute $L$? 
(The characteristic polynomial of any $\tilde{L}$  with $D/D\tilde{L} \cong N_-$ is similar, just delete the root $2^s$.) 

\item Apart from denominators or automorphisms, in what other ways can we find submodules?

\item \label{item3} The {\em zero-sum submodule}.  Let $M_0$ consist of those $h + \Delta_k(\Omega) \in M$ for which $\sum_k h$  is zero for sufficiently large $n$. 
       In our example $N_-  \subseteq M_0$,
       but in general, how to decide if $M_0$ is zero or not? How to find elements?
   
       Let $L_{\rm min}$ := MinimalRecurrence($\sum_k \ h$).  If $L \neq L_{\rm min}$ then we found a non-zero submodule
              $D \hspace{0.5pt} L_{\rm min}(m) \subseteq M_0$, but found it too late to speed up the computation of $L$ or $L_{\rm min}$.
              Is there a way to address this, to efficiently find, and then work modulo, a submodule of $M_0$?
              
      A recurrence for $a(n)$ can  be found by ``guessing'' (making an ansatz of guessed degree and order, and then solving linear equations). But the recurrence
      could be large, in which case, would it be more efficient to apply such a strategy to find elements of $M_0$ instead?

\item What is the best way to automatically find and use submodules for  hypergeometric, hyperexponential, or $D$-finite \cite{Dfinite1, Dfinite2, Dfinite3} telescoping?

\item \label{q5} If ${\rm Ann}_D(m, M/N) = {\rm Ann}_D(m, M)$, in other words, the right-factor
turns out to be all of $L$, can one still save CPU time by working
modulo $N$?
\end{enumerate}

\section{Brief comparison with prior results}
The core of current creative telescoping algorithms is reduction modulo $\Delta_k( \Omega )$.
In this paper we propose that
if we can find a nonzero submodule $N$ that does not contain the image of $H$,
to compute modulo $N$ as well, in order to compute factors of $L$ rather than computing $L$ in expanded form.
This saves CPU time, and a compact factored form is more useful, say for finding~\cite{J} closed form solutions.

The submodule approach has not only computational benefits, but theoretical ones too.
It helps explain (see $N_-$ in Section~\ref{sym} and $M_0$ in Section~\ref{FutureResearch}) why a minimal telescoper $L$ need not be a minimal
recurrence. More generally, it explains the factorization structure of $L$ from the submodule structure of $M$.

\begin{itemize}
\item Replacing $H(n,k)$ with $H(n,k) + H(n,n-k)$  to reduce non-minimality has been done before, see for instance \cite{Symmetrizing1, Symmetrizing2}
where this is was called {\em symmetrizing}.   The map $\phi_+$ in Section~\ref{sym} does the same, but Section~\ref{sym} also {\em explained why} $L$
was non-minimal.  A telescoper annihilates $m$ in $M$ instead of in $M/M_0$. 
Our approach is more general; $\phi$ is not the only useful automorphism, see Section~\ref{more}.

\item  The main prior result is that in the differential case, the paper \cite{pretelescoper} also gives a right-factor (called prescoper) based on the denominator.
\end{itemize}

Our hope for the submodule approach is that 
not only these results, but also others (Section~\ref{FutureResearch}) will
arise naturally in this framework.

\end{document}